\documentclass[a4paper,twocolumn,11pt,accepted=2024-04-24]{quantumarticle}
\pdfoutput=1

\usepackage{tikz}
\usepackage{authblk}
\usepackage{geometry}
\usepackage[english]{babel}
\usepackage[utf8]{inputenc}
\usepackage[T1]{fontenc}
\usepackage{indentfirst}
\usepackage{amsmath}
\usepackage{amssymb}
\usepackage{amsthm}
\usepackage{proof}
\usepackage{eufrak}
\usepackage[font=small,labelfont=bf]{caption}
\usepackage{graphics}

\allowhyphens

\newcommand{\fA}{\mathcal{A}}
\newcommand{\fB}{\mathcal{B}}
\newcommand{\fC}{\mathcal{C}}

\newcommand{\FF}{\mathbb{F}}

\newcommand{\Ka}{\mathcal{K}}

\newcommand{\HH}{\mathcal{H}}
\newcommand{\field}{\mathbb{Z}}
\newcommand{\complex}{\mathbb{C}}
\newcommand{\egesz}{\mathbb{Z}}

\makeatletter
\g@addto@macro\bfseries{\boldmath}
\makeatother

\renewcommand{\ge}{\geqslant}
\renewcommand{\le}{\leqslant}

\newtheorem{thm}{Theorem}
\newtheorem{biz}{Proof}
\newtheorem{example}{Example}

\newcommand{\ket}[1]{{\left|#1\right\rangle}}
\newcommand{\bra}[1]{{\left\langle #1\right|}}

\newcommand{\ZZ}{\mathcal{Z}}

\newcommand\backD{\lhd}
\newcommand\forwD{\rhd}
\newcommand\eref[1]{$(\ref{#1})$}


\usepackage{ifpdf}

\ifpdf
\usepackage{epstopdf}
\usepackage[pdftex,colorlinks,urlcolor=blue,citecolor=blue,linkcolor=blue]{hyperref}
\else
\usepackage[hypertex,colorlinks,urlcolor=blue,citecolor=blue,linkcolor=blue]{hyperref}
\fi
\begin{document}
\numberwithin{equation}{section}

\title{Tensor network decompositions for absolutely maximally entangled states}
\author{Bal\'azs Pozsgay}
\affiliation{MTA-ELTE ``Momentum'' Integrable Quantum Dynamics Research Group, Department of Theoretical Physics,
  ELTE Eötvös Loránd University, Hungary}

\author{Ian M. Wanless}
\affiliation{School of Mathematics, Monash University, Australia}

\maketitle

\begin{abstract}
  Absolutely maximally entangled (AME) states of $k$ qudits
  are quantum states that have maximal entanglement for all possible
  bipartitions of the sites/parties, and they can also be interpreted
  as perfect tensors.  We consider the problem of whether such states
  can be decomposed into a tensor network with a small number of
  tensors, such that all physical and all auxiliary spaces have the
  same dimension $D$.  We find that certain AME states with $k=6$ can
  be decomposed into a network with only three 4-leg tensors; we
  provide concrete solutions for local dimension $D=5$ and higher. Our
  result implies that certain AME states with six parties can be
  created with only three two-site unitaries from a product state of
  three Bell pairs, or equivalently, with six two-site unitaries
  acting on a product state on six qudits. We also consider the
  problem for $k=8$, where we find similar tensor network
  decompositions with six 4-leg tensors.
\end{abstract}

\section{Introduction}

Tensor networks play an important role in many branches of theoretical
physics \cite{tensor-network-review}.  They were originally developed
in the study of quantum many body systems
\cite{mps-intro1,mps-intro-2020}, where they can effectively simulate
quantum states and operators with low entanglement.  Ground states of
gapped Hamiltonians have area law entanglement, therefore physical
properties of such states can be computed with tensor networks.  In
one dimensional systems these tensor networks are called Matrix
Product States (MPS) and Matrix Product Operators (MPO).

In this work we investigate a question which is in a certain sense
opposite to the typical scenario: We ask whether tensor networks can
produce \emph{states with maximal entanglement} in certain selected
situations. There are a number of motivations to study this question,
which we explain below.

We are interested in special states of finite quantum systems, which
are called Absolutely Maximally Entangled states (AME). We consider
homogeneous cases, where the physical system is composed of $k$ sites,
and each site (or spin) is described by a $D$-dimensional Hilbert
space. A state in the tensor product Hilbert space is absolutely
maximally entangled if it has maximal entanglement for all possible
bipartitions
\cite{AME-1,AME-Helwig2,AME-Helwig3,AMEcomb1,four-AME}. AME's are
intimately related to quantum error correcting codes
\cite{AME-q-error-corr}. They can be seen as a quantum mechanical
version of Orthogonal Arrays (OA's) \cite{AMEcomb1,AMEcomb2,AMEcomb3}
known from combinatorial design theory in mathematics
\cite{OA-book}. AME's can also be interpreted as ``perfect tensors''
\cite{chaos-qchann}.

There are two central questions regarding AME's. First, for which
$D,k$ do they exist, and how many inequivalent states or families of
states can exist for a given $D,k$? Second, given a known AME, how can
we actually construct it, either in an analogue computer or in a real
world quantum computer?

Regarding the first question only partial results are known, and they
are summarized in the online Table \cite{AME-list} (see also
\cite{AME-bounds}). A famous example is the quantum version of the
problem of Euler's 36 officers, which concerns finding an AME with
$D=6$ and $k=4$. This problem has recently been solved in
\cite{euler36}, see also \cite{euler36-explanation}. The uniqueness of
the AME for $k=4$ and $D\ge 3$ was investigated recently in
\cite{arul-uniqueness}. In this work we do not provide completely new
AME's, but we contribute to the understanding of their internal
structure, by developing tensor network decompositions in certain
selected cases.

This means that we contribute to the second question above: we
contribute to understanding the structure of certain selected AME's
by finding tensor network decompositions for them. These tensor
networks can be interpreted alternatively as quantum circuits which
create these AME's from uncorrelated states. Similar questions were
considered in the works
\cite{AME-graph-2,quantum-circuits-for-AME,AMEcomb2}. Very often AME's
are constructed as ``graph states'', and such a construction can be
seen as the action of a quantum circuit on initial product sates
\cite{AME-graph-2}. Our methods are closely related, but different in
certain technical details. Most importantly, we manage to create
selected AME's with a small number of two-site gates. The number of
the gates applied is smaller than in the case of the graphs states
considered in \cite{AME-graph-2,quantum-circuits-for-AME}.

It is known that in large systems almost all states are close to
maximally entangled \cite{page-entangl}, but their actual
constructions require an exponentially large number of quantum
gates. (see also \cite{linear-growth-of-circuit-complexity}). In
contrast, we find selected states in smaller systems, which have
maximal entanglement, but which can be created with few two-site
gates.

The idea to use tensor networks to create larger objects from smaller
building blocks also appeared recently in the study of quantum error
correcting codes, see \cite{tensor-network-codes-1} and later
\cite{tensor-network-codes-2,tensor-network-codes-3}. A few years ago
AME's also appeared as building blocks of the so-called ``holographic
error correcting codes'' \cite{ads-code-1}, which can be seen as
discrete models of the AdS/CFT duality
\cite{holocode-review,holocode-review-jahn-eisert}. Our work can
contribute to this line of research, by showing that in certain cases
the fundamental multi-leg tensors in such codes can actually be
decomposed into a network of smaller tensors.

The idea of such decompositions actually appeared much earlier,
although in a slightly different context. Instead of an AME one can
also consider states which have maximal entanglement for certain
selected bipartitions, such that the choice for the allowed
bipartitions is dictated by geometrical principles. This leads
to the so-called planar maximally entangled states, also known as
perfect tangles, or block perfect tensors
\cite{perfect-tangles,planar-AME,block-perfect-tensor,planar-OA}. If
the number $k$ of constituents is even, then such states can be
interpreted as operators that act unitarily in multiple directions in a
planar arrangement of the sites. In the simplest case of $k=4$ one
obtains the so-called dual unitary operators \cite{dual-unitary-3},
which appeared earlier in pure mathematics as ``biunitary operators''
\cite{jones-planar,biunitary-permutations}, and in the holographic
code literature as ``doubly unitary operators''
\cite{Evenbly-HyMera,matt-holocode-1}.
For a treatment of bi-unitarity and its applications to quantum
information theory and quantum combinatorial objects, see
\cite{biunitary-qinf1,biunitary-qinf3}.  Generalization of these
objects to more constituents and also higher dimensional lattices have
been treated in multiple works
\cite{triunitary,ternary-unitary,prosen-mikado,huse-crystalline-circuits,sajat-multi}. In
certain cases the relevant multi-leg tensors admit a factorization
(tensor network decomposition) using 4-leg tensors. The main idea goes
back to Jones (see \cite{perfect-tangles}), but it was independently
rediscovered in various different situations
\cite{triunitary,ternary-unitary,huse-crystalline-circuits}. Our work
can be seen as a further step along this direction, with the
speciality that we consider the decomposition of actual AME's.

We should also note the recent work \cite{platonic-entanglement}, which considered the entanglement pattern of a tensor
network in the geometry of a dodecahedron. The constituting tensors were AME's and a certain local projector, and the
resulting quantum state has close to maximal entanglement, but it is not an actual AME. In our work we focus on AME's,
and we do not investigate states with big, but not maximal entanglement.

\bigskip

We focus on the cases $k=6$ and $k=8$. Our main idea is to use perfect
4-leg tensors in the decomposition. This immediately guarantees
maximal entanglement for many bipartitions of the resulting
state. Afterwards, requiring maximal entanglement for the remaining
bipartitions will pose conditions for the constituent tensors.  In our
concrete computations we will work with tensors related to linear maps
over finite fields. For $k=6$ we find decomposable AME's for selected
local dimensions $D\ge 5$.

\section{Definitions}

We consider finite Hilbert spaces of $k$ parties (or spins), each of
them carrying a Hilbert space of dimension $D\ge 2$. The full Hilbert
space is thus the tensor product
\begin{equation*}
  \mathcal{H}^{(k)}=\otimes_{j=1}^k \complex^D.
\end{equation*}
We will denote the set of sites by $S=\{1, 2, \dots, k\}$.

\subsection{Maximal entanglement}

For a given pure state $\ket{\Psi}\in\mathcal{H}$, the density matrix is
given by $\rho=\ket{\Psi}\bra{\Psi}$. For a bipartition $S=A\cup B$
with $B=S\setminus A$, the reduced density matrix of subsystem $A$ is
defined as the partial trace over subsystem $B$:
\begin{equation*}
  \rho_A=\text{Tr}_B(\rho).
\end{equation*}
The von Neumann entanglement is then defined as
\begin{equation*}
  s=-\text{Tr} \left(\rho_A \log(\rho_A)\right).
\end{equation*}
The entanglement is symmetric with respect to the exchange $A\leftrightarrow B$.

The entanglement is zero if $\ket{\Psi}$ is a product state of the
form $\ket{\Psi_A}\otimes\ket{\Psi_B}$, where $\ket{\Psi_{A,B}}$ are
states of the Hilbert spaces of the two subsystems $A$ and $B$,
respectively. Let us assume that $|A|\le |B|$. The maximal value of
the entanglement is reached when $\rho_A$ is proportional to the
identity matrix. More concretely
\begin{equation}
  \label{rhoa}
\rho_A=1/D^a, 
\end{equation}
where $a=|A|$, leading to the maximal value $s=a\log(D)$. In such a
case we say that the state $\ket{\Psi}$ is maximally entangled for the
given bipartition.

A state $\ket{\Psi}$ is said to be $\ell$-uniform, for some
positive integer $\ell$, if it has maximal entanglement for all
bipartitions with $|A|=\ell$. If a state is $\ell$-uniform then it is
also $\ell'$-uniform for any $0<\ell'<\ell$.
A state $\ket{\Psi}$ is absolutely maximally entangled (AME state), if
it is $\ell$-uniform with $\ell=\lfloor k/2\rfloor$, where
$\lfloor.\rfloor$ denotes the integer part.

Of particular interest are the cases when $k$ is even. In such a case
the state can be interpreted as a ``multiunitary matrix'' via a
state-operator correspondence \cite{AMEcomb2}. Let us spell out this
connection, by working in a concrete basis. The tensor components of
the vector $\ket{\Psi}$ can be denoted as
\begin{equation*}
  \Psi_{a_1,a_2,\dots,a_k}.
\end{equation*}
Let us choose a bipartition, where $A=\{1, 2, \dots, k/2\}$. Then we
construct a linear operator $U$ acting on the $k/2$-fold tensor
products by assigning the matrix elements as
\begin{equation}
  \label{Upsi}
  U_{a_1,a_2,\dots,a_{k/2}}^{b_1,b_2,\dots, b_{k/2}}=D^{k/4}\Psi_{a_1,a_2,\dots,a_{k/2},b_1,b_2,\dots, b_{k/2}}.
\end{equation}
Then it is immediately seen that the condition \eqref{rhoa} of
maximal entanglement is equivalent to unitarity of $U$:
\begin{equation*}
  U^\dagger U=1.
\end{equation*}

Different bipartitions into two equal subsets lead to different
unitary operators, acting from one half of the system to the other
half. However, all of the different $U$ matrices can be obtained from
each other by reshuffling their indices, corresponding to the
permutation of sites in the original vector $\ket{\Psi}$.

\subsection{4-leg tensors}

The smallest non-trivial case with an even number of tensor legs is
$k=4$. If a state $\ket{\Psi}\in \HH^{(4)}$ is an AME, then via
relations \eqref{Upsi} one obtains unitary operators that act on two
sites. It is useful to discuss this case in more detail.

\begin{figure}[t]
  \centering
  \begin{tikzpicture}
    \draw [very thick] (0,0) rectangle ++(1,1);
    \draw [thick]  (0,0) to (-0.2,-0.2);
    \draw [thick]  (1,0) to (1.2,-0.2);
    \draw [thick]  (0,1) to (-0.2,1.2);
        \draw [thick]  (1,1) to (1.2,1.2);
        \node at (-0.35,-0.35) {$1$};
        \node at (1.35,-0.35) {$2$};
        \node at (-0.35,1.35) {$4$};
        \node at (1.35,1.35) {$3$};
\node at (0.5,0.5) {$U$};
      \end{tikzpicture}
      \caption{\label{fig:du}Graphical depiction of a 4-leg
        tensor. It can be interpreted as an element of the four-fold
        tensor product space $H^{(4)}$, or alternatively, as a linear
        operator acting on the product space of two sites. }
     
\end{figure}
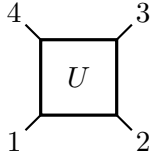

A pictorial representation of such a state/operator is given in
Fig.~\ref{fig:du}. Here the numbers stand for the spaces 1-4, which
are arranged in this representation anti-clockwise.

There are 3 different bipartitions of the set of 4 sites into two
equal subsets. A state $\ket{\Psi}$ is an AME if it has maximal
entanglement for each of these 3 bipartitions. Correspondingly, one
can obtain three unitary two-site operators using the state-operator
correspondence. First of all, we obtain a two-site operator $U$ given
by the elements
\begin{equation*}
  U_{a_1a_2}^{a_3a_4}=D\ket{\Psi_{a_1a_2a_3a_4}}.
\end{equation*}
This operator is seen in the diagram as acting from the pair of sites
$(1,2)$ to the pair $(3,4)$, thus in the vertical direction upwards.

Alternatively, we can also construct the operator $U^t$ via the correspondence
\begin{equation*}
  (U^t)_{a_1a_4}^{a_3a_2}=D\ket{\Psi_{a_1a_2a_3a_4}}.
\end{equation*}
This operator is the partial transpose of $U$ with respect to the
second space. The operator acts from the pair of sites $(1,4)$ to the
pair of sites $(3,2)$, which in the diagram is seen as acting
horizontally, from left to right.

Finally, we can construct the operator $U^R$ via
\begin{equation*}
   (U^R)_{a_1a_3}^{a_2a_4}=D\ket{\Psi_{a_1a_2a_3a_4}}.
\end{equation*}
This operator acts from the pair of sites $(1,3)$ to the pair of sites
$(2,4)$, thus on the diagram it is seen to act from one diagonal to the
other diagonal.

The state $\ket{\Psi}$ is an AME, if all three
operators $U, U^t, U^R$ are unitary.  The \emph{dual unitary}
operators are those where $U$ and $U^t$ are unitary, but $U^R$ is
arbitrary. In such a case we observe unitary action along directions
which are natural from a geometric point of view, once the sites are
arranged in the plane as the vertices of a square as in
Fig.~\ref{fig:du}. Dual unitarity is then a very natural condition in
multiple different situations, where such tensors are arranged in a
planar geometry. The two main example are holographic error correcting
codes \cite{Evenbly-HyMera}, and one dimensional quantum circuits
(where the extra dimension is given by the time coordinate)
\cite{dual-unitary-3}.

\section{Tensor network decompositions}

A tensor network is a particular way to construct states and operators
of larger Hilbert spaces using states and operators (tensors) of
smaller spaces. The idea is to take a product of a number of
fundamental tensors (each of them with a small number of indices) and
to contract some of the indices to obtain states and operators
corresponding to the larger Hilbert space. The contractions can be
visualized by the Penrose graphical notation, where the fundamental
tensors are denoted by boxes, their indices with lines (wires), and
contraction of indices is denoted by joining the lines. The remaining
free lines are then interpreted as the indices of the final object
(state or operator).

In this work we consider tensor network decompositions of specific
vectors. Our main examples will be vectors of the six-fold tensor
product $\mathcal{H}^{(6)}$, but we also consider vectors from
$\mathcal{H}^{(8)}$. Our goal is to develop tensor networks where the
fundamental tensor is an element of $\mathcal{H}^{(4)}$. Then the
graphical depiction of the network will take the form of a 4-valent
graph, with 6 or 8 outer legs, respectively.

\subsection{Decompositions for $k=6$}

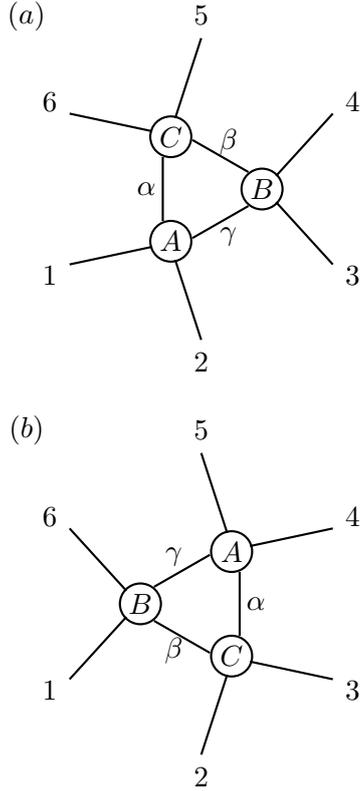
\begin{figure}[h!]
  \centering

  \begin{tikzpicture}
    \node at (135:3.25) {$(b)$};
    
    \draw[thick] (50:1) -- (30:2);
    \draw[thick] (70:1) -- (90:2);
    \draw[thick] (170:1) -- (150:2);
    \draw[thick] (190:1) -- (210:2);
    \draw[thick] (290:1) -- (270:2);
    \draw[thick] (310:1) -- (330:2);

    \draw[thick] (60:0.8) circle (0.27);
    \draw[thick] (180:0.8) circle (0.27);
    \draw[thick] (300:0.8) circle (0.27);

    \node at (60:0.8) {$A$};
    \node at (180:0.8) {$B$};
    \node at (300:0.8) {$C$};

    \node at (120:0.72) {$\gamma$};
    \node at (240:0.72) {$\beta$};
    \node at (0:0.72) {$\alpha$};

    \draw[thick] (80:0.66) -- (160:0.66);
    \draw[thick] (200:0.66) -- (280:0.66);
    \draw[thick] (320:0.66) -- (40:0.66);
    
    \node at (30:2.3) {$4$};
    \node at (90:2.3) {$5$};
    \node at (150:2.3) {$6$};
    \node at (210:2.3) {$1$};
    \node at (270:2.3) {$2$};
    \node at (330:2.3) {$3$};

    \begin{scope}[xshift=0cm,yshift=5.5cm]
      \node at (135:3.25) {$(a)$};
    \draw[thick] (10:1) -- (30:2);
    \draw[thick] (110:1) -- (90:2);
    \draw[thick] (130:1) -- (150:2);
    \draw[thick] (230:1) -- (210:2);
    \draw[thick] (250:1) -- (270:2);
    \draw[thick] (350:1) -- (330:2);

    \draw[thick] (240:0.8) circle (0.27);
    \draw[thick] (120:0.8) circle (0.27);
    \draw[thick] (0:0.8) circle (0.27);

    \node at (240:0.8) {$A$};
    \node at (0:0.8) {$B$};
    \node at (120:0.8) {$C$};

    \node at (300:0.72) {$\gamma$};
    \node at (60:0.72) {$\beta$};
    \node at (180:0.72) {$\alpha$};

    \draw[thick] (260:0.66) -- (340:0.66);
    \draw[thick] (20:0.66) -- (100:0.66);
    \draw[thick] (140:0.66) -- (220:0.66);
    
    \node at (30:2.3) {$4$};
    \node at (90:2.3) {$5$};
    \node at (150:2.3) {$6$};
    \node at (210:2.3) {$1$};
    \node at (270:2.3) {$2$};
    \node at (330:2.3) {$3$};
    \end{scope}

  \end{tikzpicture}
  
  \caption{\label{fig:deco1}Tensor networks for $k=6$.\\ (a)
    $\forwD$-decomposition, (b) $\backD$-decomposition.\\ Here
    $\alpha, \beta, \gamma$ stand for auxiliary spaces that are of the
    same dimension $D$ as the physical spaces.}
\end{figure}

In the case of $k=6$ we are looking for a decomposition into three
fundamental tensors, in the way that it is depicted in Figure
\ref{fig:deco1}. Here each circle represents a tensor with 4 legs, and
we allow the tensors to be different. Therefore, they are denoted as
$A, B, C$.

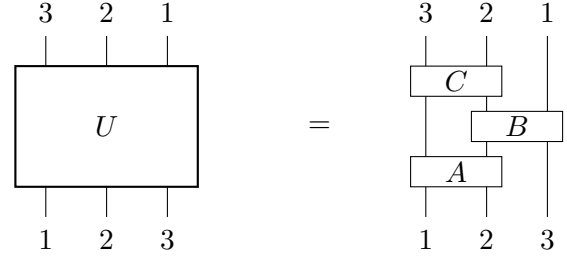
\begin{figure}[h]
  \centering
  \begin{tikzpicture}
    \draw[thick] (0,0) rectangle (2.4,1.6);
    \draw (0.4,-0.4) -- (0.4,0);
    \draw (1.2,-0.4) -- (1.2,0);
    \draw (2,-0.4) -- (2,0);
    \draw (0.4,1.6) -- (0.4,2);
    \draw (1.2,1.6) -- (1.2,2);
    \draw (2,1.6) -- (2,2);
    \node at (4,0.8) {$=$};

       \draw (7,-0.4) -- (7,0.6);
    \draw (6.2,-0.4) -- (6.2,0);
    \draw (5.4,-0.4) -- (5.4,0);
    \draw (7,1) -- (7,2);
    \draw (6.2,1.6) -- (6.2,2);
    \draw (5.4,1.6) -- (5.4,2);
    \draw (6.2,0.4) -- (6.2,0.6);
     \draw (6.2,1) -- (6.2,1.2);
\draw (5.4,0.4) -- (5.4,1.2); 
     
       \draw (5.2,1.6) rectangle (6.4,1.2);
    \draw (5.2,0.4) rectangle (6.4,0);
    \draw (6,0.6) rectangle (7.2,1);

    \node at (0.4,-0.7) {$1$};
    \node at (1.2,-0.7) {$2$};
    \node at (2,-0.7) {$3$};
       \node at (0.4,2.3) {$3$};
    \node at (1.2,2.3) {$2$};
    \node at (2,2.3) {$1$};

      \node at (5.4,-0.7) {$1$};
    \node at (6.2,-0.7) {$2$};
    \node at (7,-0.7) {$3$};
       \node at (5.4,2.3) {$3$};
    \node at (6.2,2.3) {$2$};
    \node at (7,2.3) {$1$};

    \node at (1.2,0.8) {$U$};
    \node at (5.8,1.4) {$C$};
    \node at (5.8,0.2) {$A$};
       \node at (6.6,0.8) {$B$};
  \end{tikzpicture}
  \caption{\label{fig:deco2}Decomposition of unitary operators, which
 corresponds to the $\forwD$-decomposition from
    Figure \ref{fig:deco1}. It is understood that the operator $U$
    acts from the bottom to the top. We identified the outgoing spaces
    $1, 2, 3$ of the unitary operator $U$ with the sites $4, 5, 6$ of
    the state $\ket{\Psi}$, in the given order.}
  
\end{figure}

This tensor network can be formalized most easily once we perform the
state-operator correspondence as explained in the previous
section. Then we obtain the relation
\begin{equation}
  \label{fact1}
  U^{(123)}=C^{(23)}B^{(13)}A^{(12)}.
\end{equation}
Now $U$ is a three site-operator, and $A, B$ and $C$ are two-site operators. The upper indices show the spaces on which
the operators act. 
Note that we paired the spaces 4, 5, and 6 of the original vector with
spaces 1, 2, and 3, in this order. For the graphical interpretation of
this factorization see Fig. \ref{fig:deco2}. We call this the
$\forwD$-decomposition, due to the graphical picture of the resulting tensor
network.

Note that in the picture the incoming and outgoing spaces are attached
to ``legs'' which are opposite to each other. This is a convention
that we applied also in our previous work \cite{sajat-multi}. This
convention is convenient, because in this convention the simple choice
$U=I$ with $I$ being the identity matrix satisfies the constraints of
dual unitarity and multi-directional unitarity in various cases
\cite{sajat-multi}.

We also consider the same type of decomposition, but with a different
ordering of the tensors, corresponding to the factorization
\begin{equation*}
%  \label{fact3}
  U^{(123)}=
A^{(12)}B^{(13)}C^{(23)}.
\end{equation*}
We call this the $\backD$-decomposition.  Clearly, such a decomposition
can be obtained from \eqref{fact1} simply by a permutation of the
sites $1\leftrightarrow 3$, which can be seen in the graphical
representation as a space reflection combined with a time reflection. However, it is useful to keep both
definitions at the same time. A given unitary $U^{(123)}$ might be
factorizable only in one way, or perhaps both ways.

If a factorization is possible both ways, then the two-site operators
are typically different. In those cases when the factorization is
possible both ways, with the same two-site operators, one
obtains the celebrated Yang-Baxter relation with discrete parameters:
\begin{equation}
  \label{YB}
  A^{(12)}B^{(13)}C^{(23)}=
   C^{(23)}B^{(13)}A^{(12)}.
\end{equation}
This equation is the central algebraic ingredient in the theory of
integrable models, with connections to multiple areas of pure
mathematics.

In this work we are interested in AME's. It is then very
natural to ask, whether \eqref{YB} has solutions which are perfect
tensors. To our knowledge, this question has not yet been
investigated in the literature. On the other hand, it is well known
that there are dual-unitary solutions (see for example
\cite{sajat-superintegrable}, with relations to Yang-Baxter maps).
In Example~$\ref{eg:YB}$ below, we give a solution of \eqref{YB} with
perfect tensors, such that the product is also a perfect tensor.

\subsection{Decompositions with perfect tensors}

\label{decper}

So far we did not pose any constraints for the three tensors $A, B,
C$. However, it is a very natural idea (though, as we will see later,
not absolutely necessary) that they should be chosen as perfect
tensors, in order to reach perfection of the composite object.

In this subsection we show that if $A, B, C$ are perfect tensors, then
this guarantees maximal entanglement of the six site state for
some, but not all, bipartitions. Then in the remainder of the
work we will focus on specific choices for $A, B, C$ such
that the remaining conditions of maximal entanglement can also be
satisfied.

For six sites there are 10 bipartitions, and each of them has to be
treated. We proceed by selecting the bipartitions based on a geometric
arrangement according to Figure \ref{fig:deco1}: we associate the six
sites with vertices of a regular hexagon.
We say that a quantum state on six sites (more precisely, the
corresponding operator $U$) is \emph{hexagonal unitary} or
\emph{tri-unitary}, if the state is such that it has maximal entanglement
for the three bipartitions that are obtained by cutting the hexagon
into two pieces with those symmetry axes that are disjunct from the
vertices \cite{triunitary,sajat-multi}.

\begin{thm}
 If $A, B, C$ are dual unitary operators, then the resulting six-site
 state is hexagonal unitary\footnote{This theorem is a special case of
   a general statement about similar planar arrangements, first
   published in \cite{perfect-tangles}. It is claimed in that article
   that the idea stems from Vaughan Jones. Concrete cases appeared in
   many works afterwards, often independently discovered, see for
   example \cite{huse-crystalline-circuits}.}.
 This holds for both the $\forwD$- and the $\backD$-decomposition.
\end{thm}

\begin{biz}
  Instead of a formal proof we provide a ``graphical argument'',
  using Figure~$\ref{fig:deco1}$. It is enough to check that one obtains
  a unitary operator when acting from the triplet $(1,2,3)$ to
  $(4,5,6)$, and similarly from $(2,3,4)$ to $(5,6,1)$ and also from
  $(6,1,2)$ to $(3,4,5)$. Note that in each case the resulting three
  site operator can be written as a product of three two-site
  operators. Depending on the choice of the triplets one needs to work
  with the original operators $A, B, C$ or their reshuffled versions
  $A^t, B^t, C^t$. Dual unitarity means that all six operators are
  unitary. This implies unitarity of the composite operator for all
  three choices.
\end{biz}

Let us now consider the case when $A, B, C$ are all perfect
tensors. In this case it is guaranteed that we find maximal
entanglement for three more bipartitions.  In the remainder of this
section we focus on the $\forwD$-decomposition; the
$\backD$-decomposition can be treated analogously.

\begin{thm}
Consider the $\forwD$-decomposition. If $A, B, C$ are all perfect tensors,
then the resulting six site state has maximal entanglement for the
following three bipartitions:
\begin{equation*}
\begin{aligned}
%    \begin{split}
    &(1,2,4)\cup (3,5,6),\\ %\quad
    &(1,2,5)\cup (3,4,6),\\ %\quad
    &(2,5,6)\cup (1,3,4).
%    \end{split}   
\end{aligned}
\end{equation*}
\end{thm}

\begin{biz}
  Each one of these bipartitions is such that it can be transformed
  into the previous case by permuting one of the pairs $(1,2)$,
  $(3,4)$ or $(5,6)$. For example, the bipartition $(1,2,4)\cup
  (3,5,6)$ can be transformed into $(1,2,3)\cup (4,5,6)$ by exchanging
  the pair of sites $(3,4)$. If $A, B, C$ are all perfect tensors,
  then after such a permutation they remain dual unitary, therefore we
  can apply the previous theorem to establish the unitarity of the
  resulting composite operators.
\end{biz}

An alternative proof can be given by considering a composition of
operations, acting from one subset of the sites to the complementary
set. It can be seen that there is always a sequence of operations
where we take products of two-site unitary operators, given that
$A,B,C$ are perfect.

With this we established that perfection of the constituent tensors
guarantees maximal entanglement for 6 out of the 10 bipartitions of
the state $\ket{\Psi}$ on 6 sites. The remaining 4 bipartitions are
those where we select exactly one site from each pair $(1,2), (3,4),
(5,6)$ into one part. To be concrete, the bipartitions are
\begin{equation}
\label{furcsa}
\begin{aligned}
  &(1,3,5)\cup (2,4,6),\\ %\quad
  &(1,4,5)\cup (2,3,6),\\ %\quad
  &(1,3,6)\cup (2,4,5),\\ %\quad
  &(1,4,6)\cup (2,3,5).
\end{aligned}
\end{equation}
In this case there is no way to express the operator acting from one
subset of the sites to the other subset of sites as a product of three
two-site operators. Therefore, perfection of $A, B, C$ does not
guarantee that $\ket{\Psi}$ is an AME. The unitarity conditions for the
bipartitions \eqref{furcsa} need to be checked.

As it was explained above, in this work we usually consider
decompositions into perfect tensors. In Section~\ref{s:k8} we will see
that other decompositions are sometimes possible. However, in
Section~\ref{sec:cond} we will find that in the linear case when $k=6$
that the only possible decompositions are those into perfect tensors.

At the same time, we point out that such a simple decomposition of AME
states are unlikely to exist for local dimensions $D=2$ and $D=3$. In
dimension $D=2$ there is no perfect tensor with four legs, and for
$D=3$ there is only one perfect tensor up to local unitary
transformations \cite{arul-uniqueness}.
In accordance with these simple observation, our
methods find decompositions only for bigger local dimension, starting
from $D=5$.

\subsection{Creating AME states}\label{s:AME}

Here we show that the tensor network decompositions introduced above
can be used to create the AME's from simple initial states. For
simplicity we focus on the $\forwD$-decomposition.

The task here is to start with an initial state $\ket{\Psi_0}$ with
very simple spatial structure, and then to apply a small number of
two-site unitary operators on it, in order to produce our selected AME
states. To this end, let us start with the case $k=6$ and a simple
product state
\begin{equation*}
  \ket{\Psi_0}=\ket{v}\otimes \ket{v}\otimes \ket{v}\otimes \ket{v}\otimes \ket{v}\otimes \ket{v},
\end{equation*}
where $\ket{v}$ is any selected one-site state in the local Hilbert
spaces of dimension $D$. As a first step we create a state consisting
of three Bell pairs. Let $\Ka$ be the two-site operator that acts as
\begin{equation*}
  \Ka(\ket{v}\otimes\ket{v})=\frac{1}{\sqrt{D}}\sum_a \ket{a}\otimes\ket{a},
\end{equation*}
where now $a=1,\dots,D$ labels the states in our computational
basis. We create three Bell pairs as \footnote{The creation of the Bell pairs can be seen as a special case of the
  so-called ``graph state'' construction, if the operator $\Ka$ is chosen as the so-called
controlled Z-gate (see Appendix \ref{sec:graph}). Our final formula \eqref{creating} is also related to the construction
of the graph states, but it 
involves fewer number of gates, because we do not decompose the gates $A, B$ and $C$ into the controlled Z and Fourier
gates. For more details see Appendix \ref{sec:graph}.}
\begin{equation*}
  \Ka^{(14)} \Ka^{(25)}\Ka^{(36)}\ket{\Psi_0}.
\end{equation*}
Afterwards we act with the right hand side of the factorization \eqref{fact1}, and obtain
\begin{equation}
  \label{creating}
  \ket{\Psi}=C^{(23)}B^{(13)}A^{(12)}  \Ka^{(14)} \Ka^{(25)}\Ka^{(36)}\ket{\Psi_0}.
\end{equation}
This means that if we find a desired tensor network decomposition,
then the AME can be created from a state of three Bell pairs with only
three two-site unitary operators, and from a product state with only
six two-site unitary operators.

Later, in Section \ref{s:k8}, we also consider AME states with $k=8$
parties. In that case we find tensor network decompositions with 6
four-leg tensors. In that case we need 4 additional gates to create 4
Bell pairs, therefore in total we need 10 two-site gates to create the
AME states from product states.

It is known that AME states can also be created using the so-called
``graph state'' construction, which is reviewed in Appendix
\ref{sec:graph}. There we show that the minimal number of gates
required to create the AME states is bigger in the case of the graph
states, than with our methods using the tensor network decomposition.

\section{States from classical maps}
\label{sec:classical}

In this work we consider a specific subset of states/operators, namely those
that can be obtained from classical maps. Consider the case of an AME
with a given $k$ and local dimension $D$. As explained above, for any
bipartition of the sites into two equal subsets we obtain a unitary
operator which acts from one subset of the sites to the other
subset. We say that the AME is associated with a classical map if
these unitary operators are \emph{permutation matrices} in a selected
basis. To be more concrete, let us fix a basis for the local Hilbert
spaces, and consider the bipartition $S=(1,2,\dots,k/2)\cup
(k/2+1,\dots,k)$. The basis states of the $k/2$-fold tensor product
space can be labeled by tuplets $(a_1,a_2,\dots,a_{k/2})$, where
$a_j\in X$, with $X=\{0,1,2,\dots,D\}$. We consider bijective maps $u:
X^{k/2}\to X^{k/2}$, and use the notation
\begin{equation}
  \label{ab}
  u(a_1,a_2,\dots,a_{k/2})=(b_1,b_2,\dots,b_{k/2}).
\end{equation}
We say that a state $\ket{\Psi}\in \HH^{(k)}$ is associated with the
classical map $u$ if it can be written as
\begin{equation*}
  \ket{\Psi}=\frac{1}{D^{k/4}}\sum_{a_1,\dots,a_{k/2}} \ket{a_1,\dots,a_{k/2},b_1,\dots,b_{k/2}},
\end{equation*}
where it is understood that the $b$-variables are obtained from \eqref{ab}.

According to \eqref{Upsi}, the action of the unitary operator $U$
acting between the two subsystems is then given by
\begin{equation*}
  U\ket{a_1,a_2,\dots,a_{k/2}}=\ket{b_1,b_2,\dots,b_{k/2}}.
\end{equation*}
Therefore, it is indeed a permutation matrix on a set of size $D^{k/2}$.

The state $\ket{\Psi}$ has $D^{k/2}$ components in the given basis. It
can be seen that this is the minimum number of non-zero components in
order to fulfil the unitarity condition. Therefore we say that the
state has \emph{minimal support}.

The connection between the state $\ket{\Psi}$ and the unitary matrices
has to hold for all bipartitions, and this leads us to the concept of
\emph{orthogonal arrays} (OA).
Consider an array of size $r\times c$ consisting of symbols taken from
$X$. Consider a subset of $n$ columns and the resulting subarray. We
say that this subarray consists of \emph{orthogonal} columns if every
$n$-tuple $(c_1,c_2,\dots,c_n)\in X^n$ appears exactly once among the
rows of the subarray. This notion of combinatorial orthogonality has
some similarities with that from linear algebra, but it is a different
concept.

An orthogonal array of \emph{strength} $s$ on $D$ symbols is an array with
$c$ columns and $D^s$ rows, such that every subset of $s$ columns is
orthogonal in the above sense. An example of an OA can be seen in
Table~\ref{tab:oa1}.

\begin{table}[h]
  \centering
  \begin{tabular}{|c|c|c|c|}
    \hline
    $a_1$ & $a_2$ & $b_1$ & $b_2$ \\
   \hline
    0  & 0 & 0 & 0 \\ \hline
    0  & 1 & 1 & 2 \\ \hline
    0  & 2 & 2 & 1 \\ \hline
    1 & 0 & 1 & 1 \\ \hline
    1  & 1 & 2 & 0 \\ \hline
    1  & 2 & 0 & 2 \\ \hline
    2  & 0 & 2 & 2 \\ \hline
    2  & 1 & 0 & 1 \\ \hline
    2  & 2 & 1 & 0 \\ \hline
  \end{tabular}
  \caption{\label{tab:oa1}An example of an orthogonal array. Here,
    $X=\{0,1,2\}$, and the array is of strength $2$. This means that within
    every pair of columns each ordered pair of symbols is present exactly
    once. The OA is described by a linear map
    $(b_1,b_2)=(a_1+a_2,a_1-a_2)$ over $\egesz_3$.}
  
\end{table}

In the case of an AME related to classical maps, we can compile an
array with $D^{k/2}$ rows and $k$ columns, such that each row is given by
\begin{equation*}
  (a_1,a_2,\dots,a_{k/2},b_1,b_2,\dots,b_{k/2})
\end{equation*}
with some $(a_1,a_2,\dots,a_{k/2})\in X^{k/2}$, and the $b$-variables
are obtained according to \eqref{ab}. This way we obtain an OA of
strength $k/2$. It is known that from every OA of strength $k/2$ and
$k$ columns one can construct an AME \cite{AMEcomb1,AMEcomb2,AMEcomb3}.

Let us now discuss the tensor network decomposition for the states
obtained from OA's.
%We are interested in decomposition where the
%constituent tensors also come from OA's. A priori there is no reason
%to assume that every decomposition has to be like that, but focusing
%on such cases leads to much easier computations, as we will see.
Focusing again on the case of $k=6$, we consider an AME of the form
\begin{equation*}
  \ket{\Psi}=\frac{1}{D^{3/2}}\sum_{a_1,a_2,a_3} \ket{a_1,a_2,a_{3},b_1,b_2,b_{3}},
\end{equation*}
where $a_j,b_j\in \{0, 1, 2,\dots, D-1\}$, and the $b$-variables are given by the function $u: X^3\to X^3$ as
\begin{equation*}
  (b_1,b_2,b_3)=u(a_1,a_2,a_3).
\end{equation*}
Then we intend to find a factorization analogous to \eqref{fact1}: 
\begin{equation}\label{uabc}
  u^{(123)}=\fC^{(23)}\fB^{(13)}\fA^{(12)},
  %\text{ or }  u^{(123)}=\tilde \fA^{(12)}\tilde \fB^{(13)}\tilde \fC^{(23)},
\end{equation}
where now $\fA, \fB, \fC$
%or $\tilde\fA, \tilde\fB, \tilde\fC$
are classical functions acting from $X^2$ to $X^2$. Similar to the
factorization into unitary operators, we expect that these functions
will typically correspond to AME's, and therefore describe orthogonal
arrays themselves. These smaller OA's will be of size $D^2\times 4$,
and strength $2$.  To our knowledge, this type of decomposition of
OA's into smaller OA's has not yet been considered in the mathematical
literature.

Similar to the arguments presented in Subsection \ref{decper}, we can
establish that if we construct an array via the function $u: X^3\to X^3$
as given by the product in \eqref{uabc}, such that the functions
$\fA, \fB, \fC$ describe orthogonal arrays, then in the resulting
array the orthogonality condition will hold for 12 out of the 20
subsets of size 3.

\subsection{Linear maps}

Often it is convenient to work with linear maps over finite fields.
To this end, let us assume that the set $X$ is identified with a
finite field $\FF$ of order $D$. Then for general $k$ the function
$u:X^{k/2}\to X^{k/2}$ can be represented by a matrix $G$ of size
$k/2\times k/2$ over $\FF$.  For the matrix elements we will use the
standard notation
\begin{equation*}
%  \label{Gdef}
G=\left[
  \begin{array}{ccc}
    g_{11}&g_{12}&g_{13}\\
    g_{21}&g_{22}&g_{23}\\
    g_{31}&g_{32}&g_{33}\\
  \end{array}
  \right]
\end{equation*}
and it is understood that the mapping $u: X^3\to X^3$ is performed as
\begin{equation*}
  \begin{bmatrix}
    b_1 \\ b_2 \\ b_3
  \end{bmatrix}=
  G
   \begin{bmatrix}
    a_1 \\ a_2 \\ a_3
  \end{bmatrix}.
\end{equation*}

Matrices over finite fields for which all minors are
non-vanishing are called \emph{superregular} (or in some references,
\emph{full superregular}) \cite{mds-cauchy,keri-superregular}. It is
well known that such matrices are directly related to linear MDS codes,
and they generate orthogonal arrays
\cite{mds-cauchy,keri-superregular}.
One family of superregular matrices are the \emph{Cauchy matrices}.
For a finite field $\FF$ they are given by
\begin{equation}\label{e:Cauchy}
  g_{ij}=\frac{1}{x_i-y_j},
\end{equation}
where $x_i, y_j\in \FF$, for $1\le i\le k$ and $1\le j\le k$, such that
all variables are distinct. The determinant of the Cauchy matrix
\eref{e:Cauchy} is
\begin{equation*}
  \det(G)=\frac{\prod_{i<j} (x_i-x_j)(y_j-y_i)}{\prod_{i,j}(x_i-y_j)}.
\end{equation*}
It follows that the determinant is non-zero. Every submatrix is also
a Cauchy matrix, therefore all minors are non-zero, and $G$ is indeed
superregular. A Cauchy matrix of size $k/2\times k/2$ requires a total
number of $k$ distinct elements, therefore this type of superregular
matrix can exist only if $D\ge k$.

One can argue that if we keep $k$ fixed, and then increase $D$, then
the probability that a given matrix is superregular is increased, such
that in the $D\to \infty$ limit this probability becomes 1.  From the
well known formula for the order of the general linear group, we can
deduce that the probability of any given $c\times c$ minor of a
random matrix being zero is
\[
1-\prod_{i=1}^{c}(1-D^{-i}),
\]
which is of order $1/D$. For fixed $k$ we have a fixed number of minors that
need to be checked. So, by the union bound, the probability that at least one
of them is zero is still of order $1/D$. Thus the probability of a
random matrix being superregular tends to $1$. For small values of $D$ they
are comparatively rare.
 
\subsection{Factorization with linear maps}

Let us now consider the factorization problem for $k=6$. We assume
that both the function $u: X^3\to X^3$ and also the functions $\fA,
\fB, \fC: X^2\to X^2$ are linear.  Therefore, they can be represented
by matrices of size $3\times 3$ and $2\times 2$, respectively.

Starting with the functions $\fA, \fB, \fC$, we represent them as
\begin{equation*}
  \begin{aligned}
A&=\left[\begin{array}{cc}
a_{11}&a_{12}\\
a_{21}&a_{22}\\
  \end{array}
  \right]
\\
B&=\left[\begin{array}{cc}
b_{11}&b_{12}\\
b_{21}&b_{22}\\
  \end{array}
  \right]
\\
C&=\left[\begin{array}{cc}
c_{11}&c_{12}\\
c_{21}&c_{22}\\
  \end{array}
  \right].
\end{aligned}
\end{equation*}

They can be embedded into maps acting on $X^3$ simply as
\begin{equation}
  \label{ABCemb}
  \begin{aligned}
    \tilde A&=\left[\begin{array}{ccc}
a_{11}&a_{12}& 0\\
          a_{21}&a_{22}& 0\\
          0 & 0 & 1\\
  \end{array}
  \right]
\\
\tilde B&=\left[\begin{array}{ccc}
          b_{11}&0 & b_{12} \\
          0 & 1 & 0 \\
b_{21}&0 &b_{22}\\
  \end{array}
  \right]
  \\
\tilde C&=\left[\begin{array}{ccc}
                 1& 0 & 0 \\
                 0 &c_{11}&c_{12}\\
0& c_{21}&c_{22}\\
  \end{array}
  \right].
\end{aligned}
\end{equation}

Finally, the factorization on the level of $3\times3$ matrices is performed as
\begin{equation}\label{Gfact}
 G=\tilde C\tilde B\tilde A, \qquad\text{and}\qquad G=\tilde A\tilde B\tilde C
\end{equation}
for $\forwD$-decomposition and $\backD$-decomposition, respectively. 

Then the remaining task is to find matrices $G, A, B, C$, such that
one of the decompositions in \eqref{Gfact} holds, and that all the
minors of $G$ are invertible.

It is important that the factorizations in \eqref{Gfact} are never
unique: there is some gauge freedom in the choice of the
factors. Consider the diagonal matrices
\begin{equation*}
  E_1=
  \begin{bmatrix}
    e_1 &0 &0 \\
    0 &1 &0 \\
    0 & 0 &1
  \end{bmatrix},\quad
 E_3=
  \begin{bmatrix}
    1 &0 &0 \\
    0 &1 &0 \\
    0 & 0 &e_3
  \end{bmatrix},
\end{equation*}
where $e_1, e_3\in \FF$ are nonzero. Then we can write, for example,
\begin{equation*}
  G=\tilde C E_3 (E_3)^{-1}\tilde B  E_1 (E_1)^{-1}\tilde A,
\end{equation*}
which yields the gauge transformations
\begin{equation*}
  \tilde C\to \tilde C E_3,\; \tilde B\to   (E_3)^{-1}\tilde B  E_1,\; \tilde A\to  (E_1)^{-1}\tilde A,
\end{equation*}
such that the specific form of the matrices given in \eqref{ABCemb} is
retained.  

The factorization \eqref{Gfact} is very natural for matrices over the
complex numbers. One can argue that in $GL(3,\complex)$ almost all
matrices $G$ can be factorized this way: it is easiest to see this
when every matrix is close to the identity matrix, in which case we
can immediately read off the leading terms in the matrices $A, B, C$
(up to the gauge freedom). However, in the case of 
finite fields it happens with nonvanishing probability that such a
factorization is not possible. In the next section we compute the
necessary conditions for the factorization for the case of
$k=6$. Finally, in Section \ref{sec:exampl} we provide a number of
examples, with a factorizable $G$ that also leads to an AME.

\section{Conditions of factorizability}

\label{sec:cond}

There are two possible strategies for finding solutions to our
problem. One can start with the matrices $A, B, C$, requiring that
they are superregular, and afterwards requiring that the product
matrix $G$ is also superregular. Alternatively, one can start with a
superregular $G$, and investigate whether it can be factorized in the
way we require. We choose this second path, and we find that whenever
the factorization is possible, the factors are perfect in each case.

We start with a $\forwD$-decomposition of a gate $G$ into the product
\eqref{Gfact}. We assume that $A, B, C$ all correspond to dual unitary
gates. 
Hence each element of
$\{a_{11},a_{22},b_{11},b_{22},c_{11},c_{22}\}$
must be nonzero in $\FF$.
Then, by gauge transformation we may assume $a_{11}=a_{22}=b_{22}=1$.
After some algebraic manipulations (see Appendix A), this leads to
the following general solution:
% \eref{e:g11}-\eref{e:g33}:
\begin{widetext}
\begin{align}
  a_{11}&=1\label{e:1stgensol}\\
  a_{12}&=\frac{g_{12}}{g_{11}}\\
  a_{21}&=\frac{g_{21}g_{33}-g_{23}g_{31}}{g_{22}g_{33}-g_{23}g_{32}}\\
  a_{22}&=1\\
  b_{11}&=g_{11}\\
  b_{12}&=g_{13}\\
  b_{21}&=\frac{g_{31}-c_{21}a_{21}}{g_{33}}=\frac{g_{11}(g_{22}g_{31}-g_{21}g_{32})}{g_{11}g_{22}g_{33}+g_{12}g_{23}g_{31}-g_{11}g_{23}g_{32}-g_{12}g_{21}g_{33}}\\
  b_{22}&=1\\
  c_{11}&=\frac{g_{22}-a_{12}g_{21}}{1-a_{12}a_{21}}=\frac{(g_{11}g_{22}-g_{12}g_{21})(g_{22}g_{33}-g_{23}g_{32})}{g_{11}g_{22}g_{33}+g_{12}g_{23}g_{31}-g_{11}g_{23}g_{32}-g_{12}g_{21}g_{33}}\\
  c_{12}&=g_{23}\\
  c_{21}&=\frac{g_{32}-a_{12}g_{31}}{1-a_{12}a_{21}}=\frac{(g_{11}g_{32}-g_{12}g_{31})(g_{22}g_{33}-g_{23}g_{32})}{g_{11}g_{22}g_{33}+g_{12}g_{23}g_{31}-g_{11}g_{23}g_{32}-g_{12}g_{21}g_{33}}\\
  c_{22}&=g_{33}.\label{e:lastgensol}
\end{align}
\end{widetext}
If $G$ has a $\forwD$-decomposition then it
must be given by this general solution. Moreover, it is easy to check
that this solution will work when
\begin{equation}\label{e:keycond}
  g_{11}g_{22}g_{33}+g_{12}g_{23}g_{31}-g_{11}g_{23}g_{32}-g_{12}g_{21}g_{33}\ne0
\end{equation}
in $\FF$, and not otherwise. Another notable feature of the solution
is that all variables in it are necessarily nonzero (assuming $G$ is
superregular and \eref{e:keycond} holds) and that $A,B,C$ will be
nonsingular. In other words, if $G$ represents a linear perfect tensor
with 6 legs, and it has a $\forwD$-decomposition into linear
tensors with 4 legs, then those factors will themselves be perfect.
It is an open question whether this statement is true in the nonlinear
case for $k=6$.

Applying the space reflection and interchanging $A$ with $C$, we find that
the necessary and sufficient condition for the gate $G$ to
have a $\backD$-decomposition is that
\begin{equation*}%\label{e:backDkeycond}
  g_{11}g_{22}g_{33}+g_{13}g_{21}g_{32}-g_{11}g_{23}g_{32}-g_{12}g_{21}g_{33}\ne0
\end{equation*}
in $\FF$. If $G$ has a decomposition $G=\tilde A\tilde B\tilde C$ then it will
be given by
\begin{widetext}
\begin{align*}
  a_{11}&=\frac{(g_{22}g_{33}-g_{23}g_{32})(g_{11}g_{22}-g_{12}g_{21})}{
g_{11}g_{22}g_{33}+g_{13}g_{21}g_{32}-g_{12}g_{21}g_{33}-g_{11}g_{23}g_{32}}\\
  a_{12}&=g_{21}\\
  a_{21}&=\frac{(g_{12}g_{33}-g_{13}g_{32})(g_{11}g_{22}-g_{12}g_{21})}{
g_{11}g_{22}g_{33}+g_{13}g_{21}g_{32}-g_{12}g_{21}g_{33}-g_{11}g_{23}g_{32}}\\
  a_{22}&=g_{11}\\
  b_{11}&=g_{33}\\
  b_{12}&=g_{31}\\
  b_{21}&=\frac{g_{33}(g_{13}g_{22}-g_{12}g_{23})}{g_{11}g_{22}g_{33}+g_{13}g_{21}g_{32}-g_{12}g_{21}g_{33}-g_{11}g_{23}g_{32}}\\
  b_{22}&=1\\
  c_{11}&=1\\
  c_{12}&=\frac{g_{32}}{g_{33}}\\
  c_{21}&=\frac{g_{11}g_{23}-g_{13}g_{21}}{g_{11}g_{22}-g_{12}g_{21}}\\
  c_{22}&=1.
\end{align*}
\end{widetext}

Similar to the question of superregularity, we can give a probabilistic
argument that shows that for large $D$ almost all superregular
matrices can be factorized. In this simple case, if we keep $k=3$ and
increase $D$, then the probability of not finding a factorization
decreases as~$1/D$. 

\section{Examples}\label{sec:exampl}

\subsection{$D=2$ and $D=3$}

For the finite fields $\egesz_2$ and $\egesz_3$ there are no
superregular matrices of size $3\times 3$. Indeed, there are no
orthogonal arrays of strength $3$ on $6$ columns for $D\le3$, since they
do not satisfy the Bush bound \cite{Bus52}. Therefore, our methods are
not applicable in these cases.

\subsection{$D=4$}

In the case of $D=4$ we have the finite field $GF(4)$, which is given
by the four elements $\{0,1,\omega,\omega^2\}$, together with the
algebraic relations
\begin{equation*}
  \omega+\omega=\omega^2+\omega^2=0,\qquad \omega^2=\omega+1.
\end{equation*}
It is well known \cite{Bus52}
that the following Vandermonde matrix is superregular:
\begin{equation}\label{e:omega}
G=\left[
  \begin{array}{ccc}
    1&1&1\\
    1&\omega&\omega^2\\
    1&\omega^2&\omega\\
  \end{array}
  \right]
\end{equation}
and hence produces an orthogonal array with 6 columns, 4 symbols, and
strength 3.  Therefore, it leads to an AME with $D=4$ and $k=6$. It is
known \cite{keri-superregular} that up to
equivalence relations this is the only superregular matrix for $D=4$
and $k=6$.

However, this $G$ does not have a $\forwD$-decomposition since
\begin{align*}
  &g_{11}g_{22}g_{33}+g_{12}g_{23}g_{31}-g_{11}g_{23}g_{32}-g_{12}g_{21}g_{33}\nonumber\\
&=2\omega^2-2\omega=0.
\end{align*}
It also does not have a $\backD$-decomposition since
\begin{align*}
  &g_{11}g_{22}g_{33}+g_{13}g_{21}g_{32}-g_{11}g_{23}g_{32}-g_{12}g_{21}g_{33}\nonumber\\
&=2\omega^2-2\omega=0.
\end{align*}

\subsection{$D\ge 5$}

Now we find examples for all finite fields with dimension $D\ge 5$.

\begin{example}
Consider using the matrix \eref{e:omega} over some field, with some
choice of $\omega$.  Provided $\omega\notin\{0,\pm1,-2\}$, the result
will be superregular. This can be seen by considering its minors and
its determinant; the determinant is 
\begin{equation*}
  \det(G)=-\omega (\omega-1)^2(\omega+2).
\end{equation*}
Furthermore, we find
\begin{align*}
  g&_{11}g_{22}g_{33}+g_{12}g_{23}g_{31}-g_{11}g_{23}g_{32}-g_{12}g_{21}g_{33}\nonumber\\
  =&g_{11}g_{22}g_{33}+g_{13}g_{21}g_{32}-g_{11}g_{23}g_{32}-g_{12}g_{21}g_{33}\nonumber\\
  =&2\omega^2-\omega^4-\omega=\omega(1-\omega)(\omega^2+\omega-1).
\end{align*}
In any field of order strictly larger than $5$, there will be a choice of
$\omega$ such that this quantity is nonzero. In that case we will have
a $\forwD$-decomposition and a $\backD$-decomposition of a perfect tensor.
\end{example}

\begin{example}
Let
\[
G=\left[
  \begin{array}{ccc}
    1&1&1\\
    1&2&-2\\
    1&-2&-1\\
  \end{array}
  \right].
\]
Then $G$ is a perfect tensor with a $\forwD$-decomposition over fields of
characteristic $5$, because
\begin{align*}
  &g_{11}g_{22}g_{33}+g_{12}g_{23}g_{31}-g_{11}g_{23}g_{32}-g_{12}g_{21}g_{33}\nonumber\\
  &=-2-2-4+1=-7.
\end{align*}
It also has a $\backD$-decomposition because
\begin{align*}
  &g_{11}g_{22}g_{33}+g_{13}g_{21}g_{32}-g_{11}g_{23}g_{32}-g_{12}g_{21}g_{33}\nonumber\\
  &=-7.
\end{align*}
Therefore we obtain solutions for any $D=5^n$ with $n\ge 1$. This
fills the gap left by the previous example.
\end{example}

\begin{example}\label{eg:YB}
  It is easy to write down conditions under which we can factorize a
  perfect tensor into factors that satisfy the Yang-Baxter equation
  \eqref{YB}. Using \eref{ABCemb} and equating coefficients we find
  that necessary and sufficient conditions for $\tilde A\tilde B\tilde
  C=\tilde C\tilde B\tilde A$ are that
  \begin{align*}
    a_{12} c_{11} + a_{11} b_{12} c_{21} &= b_{11} a_{12}\\
    a_{12} c_{12} + a_{11} b_{12} c_{22} &= b_{12}\\
    a_{21} b_{11} &=  a_{21} c_{11} + a_{11} b_{21} c_{12}\\ 
    a_{21} b_{12} c_{21} &= a_{12} b_{21} c_{12} \\
    a_{22} c_{12} + a_{21} b_{12} c_{22} &= b_{22} c_{12} \\
    b_{21} &= a_{21} c_{21} + a_{11} b_{21} c_{22} \\
    b_{22} c_{21} &= a_{22} c_{21}  + a_{12} b_{21}  c_{22}.
  \end{align*}
  If we assume that $a_{21}$ and $c_{12}$ are nonzero (which is true when
  $A,B,C$ are perfect), then this system of equations is equivalent to
  \begin{align*}
    a_{12}&=\frac{b_{12}}{c_{12}}(1-a_{11} c_{22})\\
    c_{21}&=\frac{b_{21}}{a_{21}}(1-a_{11} c_{22})\\
    a_{22}&=b_{22}-\frac{a_{21} b_{12} c_{22}}{c_{12}}\\
    c_{11}&=b_{11}-\frac{a_{11} b_{21} c_{12}}{a_{21}},
  \end{align*}
  with $a_{11},a_{21},b_{11},b_{12},b_{21},b_{22},c_{12},c_{22}$
  arbitrary. Adding a requirement for $A,B,C$ and $G=ABC$ to be perfect
  adds restrictions that all minors are nonzero. But these conditions 
  are possible to satisfy simultaneously.
  As a concrete example, consider again $D=5$ and the superregular matrix
  \begin{equation*}
    G=
    \begin{bmatrix}
      2 &  1 &  1 \\
      2 &  2 &  3 \\
      1 &  4 &  2
    \end{bmatrix}.
  \end{equation*}
  This can be factorized in both ways, with the $2\times 2$ matrices given by
  \begin{equation*}
    A=
    \begin{bmatrix}
      3 & 4 \\
      3 & 1  
    \end{bmatrix},\quad
    B=
    \begin{bmatrix}
      4  & 1\\
      1  & 3 
    \end{bmatrix},\quad
    C=
    \begin{bmatrix}
      3  & 1   \\
      3  & 4
    \end{bmatrix}.%\qquad 
  \end{equation*}
\end{example}

\section{Extension to $k=8$}\label{s:k8}

\begin{figure}[htb]
  \centering

  \begin{tikzpicture}

    \pgfdeclarelayer{background}
    \pgfdeclarelayer{foreground}
    \pgfsetlayers{background,main,foreground}

    \fill[white] (0:1) circle (0.33);
    \fill[white] (180:0.9) circle (0.33);
    \fill[white] (90:1.4) circle (0.33);
    \fill[white] (270:1.4) circle (0.33);
    \fill[white] (120:2.8) circle (0.33);
    \fill[white] (240:2.8) circle (0.33);

  \begin{pgfonlayer}{foreground}
  
    \draw[thick] (0:1) circle (0.33);
    \draw[thick] (180:0.9) circle (0.33);
    \draw[thick] (90:1.4) circle (0.33);
    \draw[thick] (270:1.4) circle (0.33);
    \draw[thick] (120:2.8) circle (0.33);
    \draw[thick] (240:2.8) circle (0.33);

    \node (A14) at (0:1) {$A^{14}$};
    \node (A23) at (180:0.9) {$A^{23}$};
    \node (A24) at (90:1.4) {$A^{24}$};
    \node (A13) at (270:1.4) {$A^{13}$};
    \node (A34) at (120:2.8) {$A^{34}$};
    \node (A12) at (240:2.8) {$A^{12}$};
  \end{pgfonlayer}

  \begin{pgfonlayer}{background}
    \draw[thick] (240:2.8) -- (270:1.4);
    \draw[thick] (0:1) -- (270:1.4);
    \draw[thick] (0:1) -- (90:1.4);
    \draw[thick] (240:2.8) -- (180:0.9);
    \draw[thick] (90:1.4) -- (180:0.9);
    \draw[thick] (90:1.4) -- (120:2.8);
    \draw[thick] (180:0.9) -- (120:2.8);
    \draw[thick] (180:0.9) -- (270:1.4);

    \draw[thick] (235:4.1) -- (240:2.8);
    \draw[thick] (260:3.4) -- (240:2.8);
    \draw[thick] (290:2.4) -- (270:1.4);
    \draw[thick] (335:1.9) -- (0:1);

    \draw[thick] (25:1.9) -- (0:1);
    \draw[thick] (70:2.4) -- (90:1.4);
    \draw[thick] (100:3.4) -- (120:2.8);
    \draw[thick] (125:4.1) -- (120:2.8);
    
  \end{pgfonlayer}    
    
    \node at (235:4.3) {$1$};
    \node at (263:3.55) {$2$};
    \node at (291:2.6) {$3$};
    \node at (333:2.1) {$4$};

    \node at (27:2.1) {$5$};
    \node at (69:2.6) {$6$};
    \node at (97:3.55) {$7$};
    \node at (125:4.3) {$8$};

  \end{tikzpicture}
  \caption{\label{fig:deco3}Tensor network for an AME with $k=8$
    parties. The labeling of the four-leg tensors $A^{(ab)}$ with
    $1\le a<b\le 4$ follows from the operator/state correspondence
    described in the text, see eq. \eqref{Uk8}.  }
\end{figure}
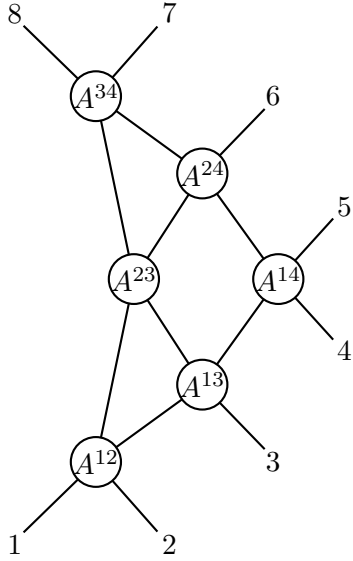

In this Section we extend the previous results to AME states with 8
parties. In this case the AME is equivalent to unitary operators $U$
that act on $\HH^{(4)}$. We are looking for a tensor network
decomposition of the AME using $4$-leg tensors, which is equivalent
to a factorization of the operator $U$ into two-site unitary
operators. There are multiple ways to perform such a
factorization. Given a selected ordering of sites, we focus on the
factorization
\begin{equation}
  \label{Uk8}
   U= A^{(34)}A^{(24)}A^{(23)}A^{(14)} A^{(13)}A^{(12)},
\end{equation}
where now the operators $A^{(jk)}$ act only on sites $j$ and $k$. This
factorization, which is illustrated in Fig.~\ref{fig:deco3}, includes
one two-site operator for each pair $(j,k)$ of sites.

Once again we only consider those AME which are related to orthogonal
arrays obtained from linear maps. Similar to the case of $k=6$, this
means that eventually we will be dealing with matrices of small sizes
over finite fields. To be more concrete, the $4$-site unitary operator
$U$ will be described by a $4\times 4$ matrix $G$.
Then we are looking for a factorization of the form
\begin{equation}\label{e:8legdecomp}
   G=
\tilde A^{(34)} \tilde A^{(24)}\tilde  A^{(23)} \tilde A^{(14)}\tilde A^{(13)}\tilde A^{(12)},
\end{equation}
where now $\tilde A^{(j,k)}$ are matrices of size $4\times 4$, such
that they are equal to the identity matrix except for the rows and
columns $j$ and $k$, where they are given by the elements of a
$2\times 2$ matrix $A^{(j,k)}$.

Direct computation and using the gauge freedom solves the
factorization as follows. Matrix elements of the factors will be
denoted as $A^{(jk)}_{lm}$, where the pairs of indices $(jk)$ stand
for the choice of the matrix, and $lm$ stand for the row and column
indices with $l,m=1,2$. We find
\begin{widetext}
\begin{align*}
A^{(12)}_{12} &= g_{12}/g_{11},&
 A^{(12)}_{21} &= m_{12}/m_{11},
\\
 A^{(13)}_{12} &= g_{13}/g_{11},
&
 A^{(13)}_{21} &= \frac{m_{13}g_{11}}{M_{12}-M_{11}},
\\
 A^{(14)}_{11} &= g_{11},
&
 A^{(14)}_{12} &= g_{14},
\\
 A^{(14)}_{21} &= {g_{11}m_{14}}/({M_{24}-M_{34}+M_{44}}),
\\
A^{(23)}_{12} &= \frac{(M_{11}-M_{12})(g_{23}(M_{24}-M_{34}+M_{44})-g_{13}g_{24}m_{14})}{N_{12}(M_{24}-M_{34}+M_{44})m_{11}},\hidewidth
\\
 A^{(23)}_{21} &= \frac{m_{11}(g_{11}g_{32}g_{44}-g_{12}g_{31}g_{44}+g_{12}g_{34}g_{41}-g_{11}g_{34}g_{42})}{N_{34}(M_{11}-M_{12})},\hidewidth
\\
 A^{(24)}_{11} &= {m_{11}N_{12}}/({M_{11}-M_{12}}),
&
 A^{(24)}_{12} &= g_{24},
\\
 A^{(24)}_{21} &= \frac{-m_{11}m_{24}N_{12}}{(M_{11}-M_{12})(M_{44}-M_{34})},
\\
 A^{(34)}_{11} &= {N_{34}m_{44}}/({M_{44}-M_{34}}),
&
 A^{(34)}_{12} &= g_{34},
\\
 A^{(34)}_{21} &= {N_{34}m_{34}}/({M_{44}-M_{34}}),
&
 A^{(34)}_{22} &= g_{44}.
\end{align*}
\end{widetext}
Here $G=[g_{ij}]$ is the final matrix, $N_{ij}=\det G[i,j|i,j]$ is the
determinant of the submatrix induced by rows and columns $i$ and $j$, and
$m_{ij}=\det G(i,j)$ is the determinant of the $3\times 3$ submatrix
of $G$ with row $i$ and column $j$ deleted, and
$M_{ij}=g_{ij}m_{ij}$. Every single matrix element that is not
included in the list above was set to unity by gauge freedom.
The conditions which are needed to prevent division by zero are:

\noindent
$M_{11}\ne M_{12}$, $M_{34}\ne M_{44}$, $M_{24}-M_{34}+M_{44}\ne0$, $N_{12}\ne0$,
$N_{34}\ne0$, $m_{11}\ne0$.\\
The last three of these conditions are automatically satisfied when
$G$ is superregular.

\subsection{Examples}

By the Bush bound \cite{Bus52} there are no orthogonal arrays on $8$
columns with strength $4$ unless $D\ge6$. Famously, there is no
orthogonal array on $4$ columns with strength $2$ with $D=6$ symbols
(since that would provide a pair of orthogonal Latin squares of order
$6$). It follows that for $D=6$ there is also no orthogonal array of
strength $3$ on $6$ columns or of strength $4$ on $8$ columns.
Hence the smallest $D$ for which a $4\times4$ superregular matrix
might occur is $D=7$. Our first two examples meets this bound.

\bigskip

\begin{example}
  We take $\FF=GF(7)$ and let $G$ be the superregular matrix
  \[
  \left[\begin{array}{cccc}
      1&1&1&1\\
      1&2&3&5\\
      1&3&2&6\\
      1&6&5&4\\
    \end{array}\right].
  \]
  By our solution it factorizes with
  \[
  \begin{array}{ll}
    A^{(12)}=\left[\begin{array}{cc}
       1 &1\\
       6 &1\end{array}\right]
    &
    A^{(13)}=\left[\begin{array}{cc}
       1 &1\\
       6 &1\end{array}\right]
    \\[3ex]
    A^{(14)}=\left[\begin{array}{cc}
       1 &1\\
       5 &1\end{array}\right]\phantom{.}
    &
    A^{(23)}=\left[\begin{array}{cc}
       1 &5\\
       4 &1\end{array}\right]
    \\[3ex]
    A^{(24)}=\left[\begin{array}{cc}
       4 &5\\
       3 &1\end{array}\right]
    &
    A^{(34)}=\left[\begin{array}{cc}
       1 &6\\
       2 &4\end{array}\right].
  \end{array}
  \]
\end{example}

\begin{example}
  We take $\FF=GF(7)$ and let $G$ be the superregular matrix
  \[
  \left[\begin{array}{cccc}
      1&1&1&1\\
      1&3&4&5\\
      1&4&5&3\\
      1&5&3&4\\
    \end{array}\right].
  \]
  By our solution it factorizes with
  \[
  \begin{array}{ll}
    A^{(12)}=\left[\begin{array}{cc}
       1 &1\\
       3 &1\end{array}\right]
    &
    A^{(13)}=\left[\begin{array}{cc}
       1 &1\\
       2 &1\end{array}\right]
    \\[3ex]
    A^{(14)}=\left[\begin{array}{cc}
       1 &1\\
       5 &1\end{array}\right]\phantom{.}
    &
    A^{(23)}=\left[\begin{array}{cc}
       1 &0\\
       0 &1\end{array}\right]
    \\[3ex]
    A^{(24)}=\left[\begin{array}{cc}
       6 &5\\
       3 &1\end{array}\right]
    &
    A^{(34)}=\left[\begin{array}{cc}
       4 &3\\
       4 &4\end{array}\right].
  \end{array}
  \]
  Notably, $A^{(23)}$ is not perfect in this case, even though $G$ is
  superregular.  Such behaviour was impossible in the $k=6$ case.
  In fact, since $A^{(23)}$ is the identity matrix, we see that this
  particular $G$ can be factorized using just five $4$-leg tensors.
\end{example}

Our next example is for $D=8$. A local state of dimension $8$ can be
understood as the tensor product space of three qubits, therefore such
a solution could be interesting quantum computation based on qubits.

\begin{example}
  Let $\FF=GF(8)$ be the field of order $8$. It can be understood as
  $\field_2[x]/\langle x^3+x+1\rangle$, or more
  formally as the algebra defined by the basis elements $1, x, x^2$
  with coefficients in $\egesz_2$ and the additional relation
    $x^3=x+1$.
Now let us take $G$ to be
  \[
  \left[\begin{array}{cccc}
  x&x^2&x^2+x&x^2+1\\
x^2+x&1&x&x^2+x+1\\
x^2+x+1&x+1&x^2+1&x^2+x\\
1&x^2+x&x^2&x+1
    \end{array}
    \right].
  \]
  This matrix is a Cauchy matrix with
  $(x_1,x_2,x_3,x_4)=(0,x^2+x,1,x^2)$ and
  $(y_1,y_2,y_3,y_4)=(x^2+1,x^2+x+1,x+1,x)$ in \eref{e:Cauchy},
  therefore all of its minors are nonzero, so we have a perfect
  tensor. Also, by our solution above it factorizes with
%\begin{widetext}
  \[
  \begin{array}{l}
    A^{(12)}=\left[\begin{array}{cc}
1&x\\
x+1&1\end{array}\right]
\\[3ex]
    A^{(13)}=\left[\begin{array}{cc}
1&x+1\\
x^2+1&1\end{array}\right]
\\[3ex]
    A^{(14)}=\left[\begin{array}{cc}
x&x^2+1\\
x^2+1&1\end{array}\right]\phantom{.}
\\[3ex]
    A^{(23)}=\left[\begin{array}{cc}
1&x^2+x\\
x^2+x&1\end{array}\right]
\\[3ex]
    A^{(24)}=\left[\begin{array}{cc}
x^2+1&x^2+x+1\\
x^2+1&1\end{array}\right]
\\[3ex]
    A^{(34)}=\left[\begin{array}{cc}
1&x^2+x\\
x^2+x&x+1\end{array}\right].
    \end{array}
  \]
%  \end{widetext}
\end{example}

\begin{example}
  We take $\FF=GF(11)$ and let $G$ be the superregular matrix
  \[
  \left[\begin{array}{cccc}
      1&1&1&1\\
      1&2&3&4\\
      1&3&2&9\\
      1&7&8&5\\
    \end{array}\right].
  \]
  By our solution it factorizes with
  \[
  \begin{array}{ll}
    A^{(12)}=\left[\begin{array}{cc}
       1 &1\\
       8 &1\end{array}\right]
    &
    A^{(13)}=\left[\begin{array}{cc}
       1 &1\\
       10 &1\end{array}\right]
    \\[3ex]
    A^{(14)}=\left[\begin{array}{cc}
       1 &1\\
       3 &1\end{array}\right]\phantom{.}
    &
    A^{(23)}=\left[\begin{array}{cc}
       1 &8\\
       0 &1\end{array}\right]
    \\[3ex]
    A^{(24)}=\left[\begin{array}{cc}
       3 &4\\
       8 &1\end{array}\right]
    &
    A^{(34)}=\left[\begin{array}{cc}
       4 &9\\
       3 &5\end{array}\right].
  \end{array}
  \]
  Again, $A^{(23)}$ is not perfect, although in this example it is not
  the identity. Note that by inspecting the general solution we can see
  that $A^{(23)}$ is the only one of the 6 factors which might contain
  a zero entry.

  Permuting the rows of $G$ we get another superregular
  matrix
  \[
  G'=\left[\begin{array}{cccc}
      1&7&8&5\\
      1&1&1&1\\
      1&2&3&4\\
      1&3&2&9\\
    \end{array}\right].
  \]
  For this $G'$ there is no decomposition of the form \eref{e:8legdecomp},
  since it satisfies $M_{34}=M_{44}=10$.
\end{example}

\section{Discussion}

In this work we obtained tensor network decompositions for selected
AME states. Our results are based on the classical linear maps.  A
drawback of our method is that we could obtain solutions only for
relatively large local Hilbert spaces. For $k=6$ we found solutions
starting from $D\ge 5$, and for $k=8$ our smallest solution is for
$D=7$. It would be interesting to develop alternative methods that
could settle the problem for smaller local dimension. It would be
interesting to establish decomposability or perhaps indecomposability
of selected known AME's with $D=2, 3, 4$.

Interestingly, we find that for large $D$ it becomes easier to satisfy
the conditions of factorizability. We gave a probabilistic argument
that keeping $k$ fixed and letting $D\to\infty$ we will find that
almost all perfect tensors built from linear maps can be factorized.
This is parallel to the statement that keeping $k$ fixed and letting
$D\to\infty$ it is always possible to find an AME \cite{AME-Helwig3},
and that in larger and larger Hilbert spaces almost all states are
close to maximally entangled \cite{page-entangl}.

Returning to small local dimensions, we remark that there are no
perfect $4$-leg tensors for $D=2$ \cite{no-qubit-AME}. Therefore we
conjectured that AME states with qubits cannot be decomposed in the
way that we formulated here. It is known that for qubits all AME
states are essentially graph states, which means that they can be
obtained by the action of a larger number of two-site unitary
operators on product states, but not with the small number of gates
that we find for higher dimensions (see the discussion in Appendix \ref{sec:graph}).

Alternatively, one can also consider a weaker set of constraints, by
requiring maximal entanglement only for a selected set of
bipartitions. As was mentioned in the Introduction, for such cases
decompositions were already presented in the literature in multiple
independent works
\cite{perfect-tangles,triunitary,ternary-unitary,huse-crystalline-circuits}. It
would be useful to develop a general theory for such
decompositions/factorizations.

\bigbreak
\bigskip\bigskip
\noindent
{\bf Acknowledgments} 
\nobreak

%\noindent
We are thankful for discussions with
Suhail Ahmad Rather,
S. Aravinda,
J\'anos Asb\'oth,
Wojciech Bruzda,
M\'at\'e Matolcsi,
Mil\'an Mosonyi,
Jamie Vicary,
P\'eter Vrana,
Mih\'aly Weiner,
Tianci Zhou,
Zolt\'an Zimbor\'as,
and Karol \.{Z}yczkowski.

%\addcontentsline{toc}{section}{References}
%\bibliography{pozsi-general} 
%\bibliographystyle{quantum}

\providecommand{\href}[2]{#2}\begingroup\raggedright\endgroup

\appendix 

\section{Details of the decomposition for $N=6$}

Here we give the derivation of \eref{e:1stgensol}--\eref{e:lastgensol}.
Equating coefficients in $G=\tilde C\tilde B\tilde A$ we get
\begin{align}
  g_{11}&=b_{11}\label{e:g11}\\
  g_{12}&=b_{11}a_{12}\label{e:g12}\\
  g_{13}&=b_{12}\label{e:g13}\\
  g_{21}&=c_{11}a_{21}+c_{12}b_{21}\label{e:g21}\\
  g_{22}&=c_{11}+c_{12}b_{21}a_{12}\label{e:g22}\\
  g_{23}&=c_{12}\label{e:g23}\\
  g_{31}&=c_{21}a_{21}+c_{22}b_{21}\label{e:g31}\\
  g_{32}&=c_{21}+c_{22}b_{21}a_{12}\label{e:g32}\\
  g_{33}&=c_{22}.\label{e:g33}
\end{align}
Since $b_{11}\ne0$, we see from \eref{e:g11} and \eref{e:g12} that
\begin{equation}\label{e:a12}
  a_{12}=\frac{g_{12}}{g_{11}}.
\end{equation}
Substituting \eref{e:g22} and \eref{e:g23} into \eref{e:g21} gives
\begin{equation}\label{e:ng21}
g_{21}=(g_{22}-g_{23}b_{21}a_{12})a_{21}+g_{23}b_{21}.
\end{equation}
Similarly, substituting \eref{e:g32} and \eref{e:g33} into \eref{e:g31} gives
\begin{equation}\label{e:ng31}
g_{31}=(g_{32}-g_{33}b_{21}a_{12})a_{21}+g_{33}b_{21}.
\end{equation}
Multiplying \eref{e:ng21} by $g_{33}$ and subtracting $g_{23}$ times
\eref{e:ng31}, we find that
\begin{equation*}%\label{e:ng2123}
g_{33}g_{21}-g_{23}g_{31}=(g_{33}g_{22}-g_{23}g_{32})a_{21}.
\end{equation*}
Now $G$ is superregular, so its order 2 minor $g_{33}g_{22}-g_{23}g_{32}$
is nonzero in $\FF$.
Hence
\begin{equation*}%\label{e:a21}
a_{21}=\frac{g_{33}g_{21}-g_{23}g_{31}}{g_{33}g_{22}-g_{23}g_{32}}.
\end{equation*}
Next, substituting \eref{e:g21} into \eref{e:g22} we find that
\[
g_{22}=c_{11}(1-a_{12}a_{21})+a_{12}g_{21}.
\]
From this we conclude that either $a_{12}a_{21}=1$ and $g_{22}=a_{12}g_{21}$,
or $a_{12}a_{21}\ne1$ and
\begin{equation}\label{e:c11}
c_{11}=\frac{g_{22}-a_{12}g_{21}}{1-a_{12}a_{21}}.
\end{equation}
Suppose that the former situation occurs. Then, by \eref{e:a12}
we get that
$g_{12}g_{21}=g_{11}g_{22}$. This says that the leading $2\times2$ minor of $G$
is 0, contradicting superregularity. Hence, \eref{e:c11} must hold.

Applying similar reasoning, we find that
\begin{equation*}%\label{e:c21}
c_{21}=\frac{g_{32}-a_{12}g_{31}}{1-a_{12}a_{21}}.
\end{equation*}
Then from \eref{e:g31} and the fact that $c_{22}\ne0$ we find that
\begin{equation*}%\label{e:b21}
b_{21}=\frac{g_{31}-c_{21}a_{21}}{g_{33}}.
\end{equation*}
Equations \eref{e:1stgensol}--\eref{e:lastgensol} now follow.

\section{Comparison with graph states}

\label{sec:graph}

It is known that the so-called graph states can be used to obtain
AME's for various values of the number of sites $k$ and local
dimension $D$ \cite{AME-graph}. Here we review this construction, and
we consider selected examples. We compare this construction of AME's
with our results. We assume again that $k$ is even, and for simplicity
we consider the case when $D$ is a prime \cite{AME-graph}; for $D$
being a prime power, see \cite{AME-graph-2}.

Let us choose a basis in the local spaces, denoted by $\ket{a}$,
$a=0,\dots,D-1$ and let us introduce the so-called controlled
$Z$-operator,
% [[SHOULD THIS BE $\ZZ$-operator?]],
acting on the tensor product $\complex^D\otimes
\complex^D$ as
\begin{equation*}
  \ZZ=\sum_{m,n=0}^{D-1} \omega^{mn}  (\ket{m}\bra{m}\otimes \ket{n}\bra{n} ),
\end{equation*}
where
\begin{equation*}
  \omega=e^{2\pi i/D}.
\end{equation*}
We extend this operator to the two-site operators $\ZZ_{i,j}$ that act
on the $N$-fold tensor product of the local spaces $\complex^D$, such
that $\ZZ_{i,j}$ acts on the site $i$ and $j$ non-trivially. Note the
exchange symmetry $i\leftrightarrow j$.

Graph states can be constructed as follows. We define an ``initial state'', which is a tensor product of identical local states:
\begin{equation*}
  \ket{\Psi_0}=\otimes_{j=1}^N \left(\frac{1}{\sqrt{D}}\sum_{a=0}^{D-1}\ket{a}\right).
\end{equation*}
To each pair of non-identical sites $(j,k)$ we associate an integer
$\ell=0, 1,\dots,D$. The graph states are then created as
\begin{equation*}
  \ket{\Psi}=\prod_{i<j} \left(\ZZ_{i,j}\right)^{\ell_{i,j}} \ket{\Psi_0}.
\end{equation*}
The $\ZZ_{i,j}$ operators commute with each other, because they are all diagonal in the selected basis. Therefore, the
ordering of the operator product is irrelevant. Such states are uniquely characterized by the numbers $\ell_{i,j}$ with
$i<j$. These numbers encode a weighted graph on $N$ vertices, hence the name ``graph states''. In the following we
denote by $L$ the matrix of size $N\times N$ obtained via $L_{i,j}=L_{j,i}=\ell_{i,j}$ for any $i<j$ and $L_{j,j}=0$ for
any $j$. 

The total number of $\ZZ$ gates applied for a selected state is
\begin{equation*}
  \sum_{i<j} \ell_{i,j}.
\end{equation*}
However, for any $\ell_{i,j}\ne 0$ one could count the operator power $\left(\ZZ_{i,j}\right)^{\ell_{i,j}}$ as a single
two-site gate. Then the number of the two-site gates applied is seen as the number of non-zero $\ell_{i,j}$. 

If one starts from the product state $\ket{\Psi_0}$, then the
application of a two-site gate $\ZZ_{i,j}$ will create ``one unit of entanglement'' (a Bell pair) in
the state between sites $i$ and $j$.  However, it is not obvious what
is the entanglement pattern in the final state, once many gates are
applied. This problem was solved in the work \cite{AME-graph}. It was
shown that the final state is an AME if the symmetric matrix $L$ satisfies the
following property:

\medskip
\noindent
{\bf Property 1:} For any bipartition of the numbers $(1, 2, \dots, k)$
into two equal sized subsets $A$ and $B$, the minor of size
$k/2\times k/2$ of $L$ corresponding to the rows taken from subset
$A$ and columns from subset $B$ is non-zero in the finite field $\FF$.

\medskip

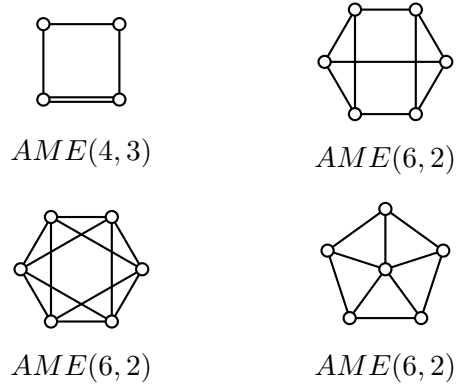
\begin{figure}[htb]
  \centering

  \begin{tikzpicture}
    \begin{scope}[xshift=-0.5cm,yshift=-0.5cm]
         \draw[thick] (0,0) -- (0,1) -- (1,1) -- (1,0);
    \draw[thick] (0,0.03) -- (1,0.03);
     \draw[thick] (0,-0.03) -- (1,-0.03);
    \draw[thick,fill=white] (0,0) circle (0.08);
    \draw[thick,fill=white] (0,1) circle (0.08);
    \draw[thick,fill=white] (1,0) circle (0.08);
    \draw[thick,fill=white] (1,1) circle (0.08);
    \node at (0.5,-0.7) {$AME(4,3)$};
      \end{scope}

    \begin{scope}[xshift=4cm]
      \draw[thick] (0:0.8) -- (60:0.8) -- (120:0.8) -- (180:0.8) -- (240:0.8) -- (300:0.8) -- (0:0.8);
     \draw[thick] (0:0.8) -- (180:0.8);
     \draw[thick] (60:0.8) -- (300:0.8);
     \draw[thick] (120:0.8) -- (240:0.8);
     
     \draw[thick,fill=white] (0:0.8) circle (0.08);
      \draw[thick,fill=white] (60:0.8) circle (0.08);
      \draw[thick,fill=white] (120:0.8) circle (0.08);
      \draw[thick,fill=white] (180:0.8) circle (0.08);
      \draw[thick,fill=white] (240:0.8) circle (0.08);
      \draw[thick,fill=white] (300:0.8) circle (0.08);
  \node at (0,-1.3) {$AME(6,2)$};      
    \end{scope}
      
    \begin{scope}[xshift=0cm,yshift=-2.75cm]
      \draw[thick] (0:0.8) -- (60:0.8) -- (120:0.8) -- (180:0.8) -- (240:0.8) -- (300:0.8) -- (0:0.8);
      \draw[thick] (0:0.8) --  (120:0.8) --  (240:0.8) -- (0:0.8);
      \draw[thick]  (60:0.8)  -- (180:0.8)  -- (300:0.8) -- (60:0.8);
      \draw[thick,fill=white] (0:0.8) circle (0.08);
      \draw[thick,fill=white] (60:0.8) circle (0.08);
      \draw[thick,fill=white] (120:0.8) circle (0.08);
      \draw[thick,fill=white] (180:0.8) circle (0.08);
      \draw[thick,fill=white] (240:0.8) circle (0.08);
      \draw[thick,fill=white] (300:0.8) circle (0.08);
  \node at (0,-1.3) {$AME(6,2)$};      
    \end{scope}

  \begin{scope}[xshift=4cm,yshift=-2.75cm]
      \draw[thick] (18:0.8) -- (90:0.8) -- (162:0.8) -- (234:0.8) -- (306:0.8) -- (18:0.8);
      \draw[thick] (0:0) -- (18:0.8);
      \draw[thick] (0:0) -- (90:0.8);
      \draw[thick] (0:0) -- (162:0.8);
      \draw[thick] (0:0) -- (234:0.8);
      \draw[thick] (0:0) -- (306:0.8);
      \draw[thick,fill=white] (0:0) circle (0.08);
      \draw[thick,fill=white] (18:0.8) circle (0.08);
      \draw[thick,fill=white] (90:0.8) circle (0.08);
      \draw[thick,fill=white] (162:0.8) circle (0.08);
      \draw[thick,fill=white] (234:0.8) circle (0.08);
      \draw[thick,fill=white] (306:0.8) circle (0.08);
  \node at (0,-1.3) {$AME(6,2)$};      
    \end{scope}
    
  \end{tikzpicture}
  
  \caption{\label{fig:graphs}Examples for weighted graphs for
    $AME(k,D)$ with different values of the number of parties $k$ and
    local dimension $D$. The number of single lines stands for the
    weight $\ell_{i,j}$ between sites $i$ and $j$. The number of
    two-site gates needed to create these AME using the controlled
    $Z$-gates is 4, 9, 10 and 12, respectively.}
  
\end{figure}

Examples for graphs for which the graph states are AME are shown in
Fig \ref{fig:graphs}.

It is important that a selected AME can have multiple representations
as a graph state. The different representations can be transformed
into each other via local unitary operations, which can be represented
also on the graphs themselves.  This problem was investigated, for
example, in \cite{graph-clifford-1,graph-clifford-2}.

We next argue that the minimal number of gates required to create an
AME via its graph state representation is $(k/2)^2$. The equivalent
mathematical problem is to find a symmetric matrix $L$ over the finite
field $\FF$, satisfying Property 1, such that the number of non-zero
elements is minimal. In any column $i$ of $L$ there cannot be $k/2$
zeros in rows other than row $i$. If there were, then choosing those
rows to form $A$ would result in a submatrix with a column of zeros,
leading to a minor that is zero.  Thus at most $k/2$ entries in column
$i$ are zero. As this is true for all $k$ columns, there are at most
$k^2/2$ zeros in $L$, corresponding to $k^2/4$ gates in the AME.
Equation~(\ref{e:GGT}) below shows one way to achieve this many zeros.

\bigskip

Now we show that the states that we treated in
Section~\ref{sec:classical}, which are obtained from classical linear
maps, are in fact equivalent to graph states of a special form. Let
us start with the formula for this type of states
\begin{equation}
  \label{cl}
  \ket{\Psi}=\frac{1}{D^{k/4}}\!\!\!\!\sum_{\atop a_1,\dots,a_{k/2}=1}^D \!\!\!\!\ket{a_1,\dots,a_{k/2},b_1,\dots,b_{k/2}},
\end{equation}
where it is now understood that for each tuplet $(a_1,a_2,\dots,a_{k/2})$ we have
\begin{equation*}
  \begin{bmatrix}
    b_1 \\ b_2 \\ \vdots \\ b_{k/2}
  \end{bmatrix}=G
    \begin{bmatrix}
    a_1 \\ a_2 \\ \vdots \\ a_{k/2}
  \end{bmatrix},
\end{equation*}
where $G$ is a matrix with size $k/2\times k/2$ over $\FF$.

Let $F$ be the Fourier matrix of size $D\times D$ with its matrix
elements given by
\begin{equation*}
  F_{ab}=\frac{1}{\sqrt{D}}\omega^{ab}.
\end{equation*}
Let us now act with the Fourier transform on the last $k/2$ spaces. Formally, we write
\begin{equation*}
  \ket{\Psi'}=(1\otimes 1\otimes\dots\otimes 1\otimes F\otimes F\otimes \dots\otimes F)\ket{\Psi},
\end{equation*}
where $\ket{\Psi}$ is a state of the form \eqref{cl}. We get
\begin{widetext}
\begin{equation*}
%  \label{cl2}
  \ket{\Psi'}=\frac{1}{D^{k/2}}\sum_{a_m,c_n=1}^D
\omega^{\sum_{j,m=1}^{k/2} G_{jm}c_j a_m}
  \ket{a_1,a_2,\dots,a_{k/2},c_1,c_2,\dots,c_{k/2}}.\\
\end{equation*}
\end{widetext}

This is a graph state with an incidence matrix of size $k\times k$  given by the block matrix
\begin{equation}\label{e:GGT}
  L=
  \begin{pmatrix}
    0 & G^T \\  G & 0
  \end{pmatrix},
\end{equation}
where each block is of size $k/2\times k/2$. If $G$ is super-regular,
then $L$ satisfies Property 1.

The special graph states of the above form are created with a total
number of $(k/2)^2$ gates from the product states $\ket{\Psi_0}$. In
the cases of $k=6$ and $k=8$ this means a total number of 9 and 16
gates. As was shown above, this is the minimal number of gates with
the graph state construction.

This is to be contrasted with our construction for the tensor
network decompositions, where we create the same states with a total
number of 6 and 10 gates, respectively (see Section~\ref{s:AME}).

\end{document}